\documentclass{ws-p8-50x6-00}

\begin{document}

\title{Theory of the Nucleon Spin--Polarizabilities~I}

\author{Ulf-G. Mei{\ss}ner}

\address{Forschungszentrum J\"ulich,
Institut f\"ur Kernphysik (Theorie)\\ D-52425 J\"ulich, 
Germany\\E-mail: Ulf-G.Meissner@fz-juelich.de}


\maketitle

\abstracts{I summarize the theoretical predictions for the
spin--dependent nucleon polarizabilities
based on chiral effective field theory approaches.}

\section{Introduction}

Low--energy Compton scattering off the nucleon is an important
probe to unravel the nonperturbative structure of QCD since the electromagnetic
interactions in the initial and final state are well understood.
In the long wavelength limit, only the charge of the target can be detected,
which is nothing but the celebrated Thomson low energy theorem.
At higher photon energies, $50<\omega<100$ MeV, the internal
structure of the system slowly becomes visible.
These nucleon structure--dependent effects in {\it unpolarized}
Compton scattering are taken into account by introducing {\it two} free
parameters into the cross-section formula, commonly denoted the
{\it electric} $(\bar\alpha)$ and {\it magnetic} $(\bar\beta)$ polarizabilities of
the nucleon in analogy to the structure dependent response functions for
light-matter
interactions in classical electrodynamics. Over the past few decades several
experiments on low energy Compton scattering off the proton have taken place,
resulting
in several extractions of the electromagnetic polarizabilities of the proton.
At present,
the commonly accepted numbers are
$\bar\alpha^{(p)} = (12.1\pm 0.8\pm 0.5)\times 10^{-4}\,{\rm fm}^3$,
$\bar\beta^{(p)} = (2.1\mp 0.8\mp 0.5)\times 10^{-4}\,{\rm fm}^3$,\cite{1}
indicating that the proton compared to its volume of $\sim 0.5\,$fm$^3$ is 
a rather stiff object.  At present, several quite different theoretical
approaches
find qualitative and quantitative explanations for these two polarizabilities,
but they also constitute one of the striking successes of chiral
perturbation theory.\cite{BKMpola,ulfco} The lowest order predictions
stem from finite loop graphs and can be expressed in terms of well--known
parameters
\begin{equation}
\bar{\alpha}_p =\bar{\alpha}_n = 10\bar{\beta}_p = 10\bar{\beta}_n =
{5 e^2 g_A^2 \over 384 \pi^2 F_\pi^2 M_\pi} = 12.4 \cdot 10^{-4} \,
{\rm fm}^3~,
\end{equation}
with $g_A = 1.26$ the axial--vector coupling measured in neutron $\beta$--deacy,
$F_\pi = 92.4\,$MeV the  pion decay constant, $M_\pi = 139.57\,$MeV the
pion mass and $e^2/4\pi = 1/137.036$ the fine structure constant. 
These numbers are in striking agreement with the experimental values,
demonstrating the importance of the pion cloud to the structure of the nucleon.
In ref.\cite{ulfco} it was further demonstrated that the fourth order corrections
to these results are modest and in particular, a novel large and
negative pion loop contribution to the magnetic polarizability was found,
leading to a substantial reduction of the sizeable positive $\Delta$ contribution.
   
Quite recently, with the advent of polarized targets and new sources with a
high
flux of polarized photons, the case of {\it polarized} Compton scattering off
the
proton $\vec{\gamma}\,\vec{p}\rightarrow\gamma p$ has come close to
experimental
feasibility. On the theoretical side it has been shown \cite{ragusa} that one
can
define {\em four} spin--dependent electromagnetic response functions
$\gamma_i,\;i=1\dots 4$, which in analogy to $\bar{\alpha}\, , \bar{\beta}$
are
commonly called the ``spin-polarizabilites'' of the proton. I remark that
one can also use a more physical basis in terms of electric and magnetic
dipole, quadrupole, $\ldots$ excitations, as discussed by Hemmert in these 
proceedings.\cite{trhproc} First studies have
been published,\cite{bab,gorchtein}
claiming that the such parameterized information on the low--energy spin
structure of the proton can really be extracted from the upcoming
double-polarization
Compton experiments. A success of this program would clearly shed new light on
our understanding of the internal dynamics of the proton and at the same time
serve
as a check on the theoretical explanations of the polarizabilities. The new
challenge to theorists will then be to explain all six of the leading
electromagnetic
response functions simultaneously. At present there  exist two
experimental
analysis that have shed some light on the magnitude of the (essentially) unknown
spin-polarizabilities $\gamma_i^{(p)}$ of the proton. First,
the LEGS group has reported~\cite{legs} a result for a linear
combination involving three of the $\gamma_i$ (for the proton), namely
\begin{equation}
\gamma_\pi^{(p)} =\gamma_1^{(p)}+\gamma_2^{(p)}+2\gamma_4^{(p)}~,
\end{equation}
whose magnitude could not be explained by theoretical approaches. 
We note that this pioneering
result was obtained from an analysis of an {\em unpolarized} Compton
experiment in the backward direction, where the spin-polarizabilities come in
as
one contribution in a whole class of subleading order nucleon structure effects
in the
differential cross-section. However, the G\"ottingen group has  performed a
similar measurement at MAMI~\cite{Wiss} 
and their value for $\gamma_\pi^{(p)}$ comes out consistent
with theoretical expectations.\footnote{I remark that as explained below, one should 
subtract the pion pole (anomaly) contribution from the true nucleon--structure
effects. This is often not done.} While such indirect determinations of some linear
combinations of the spin--polarizabilities are very valuable,
we can only reemphasize the need for the upcoming polarized
Compton scattering experiments to disentangle the contributions from the {\em four}
spin--polarizabilities.
It is therefore a challenge on the theory side to work out the complete one--loop
(that is fourth order) representation of the spin--polarizabilities
within the context of Heavy Baryon Chiral
Perturbation Theory (HBCHPT), extending previous efforts
\cite{bkmrev,hhkk,Ji,Birse}
in a significant way. These results have been reported in,\cite{GHMprl}
and a similar study can be found in.\cite{Birse2}
The active degrees of freedom in HBCHPT are
the asymptotically observable pion and nucleon fields. The various
contributions from tree and loop diagrams are organized according to
power counting rules, i.e. one expands in small momenta and pion
masses ($M_\pi$), collectively denoted by $p$.
Previously an order ${\cal O}(p^3)$ SU(2) HBCHPT calculation
\cite{bkmrev} was performed, which showed that the leading (i.e. long--range)
structure effects in the spin-polarizabilities are given by 8 different $\pi N$
loop diagrams giving rise to a $1/M_\pi^2$ behavior in the $\gamma_i$ (which is 
known since long, see~\cite{BKKM}). 
Subsequently, it was shown in a third order SU(2)
calculation,\cite{hhkk} in which the first nucleon resonance,
the $\Delta$(1232), was included as an explicit
degree of freedom,\cite{SSE} that two ($\gamma_2,\;\gamma_4$) of the four
spin--polarizabilities receive
large corrections due to $\Delta$(1232) related effects, resulting in a big
correction to the leading $1/M_\pi^2$ behavior. In that
phenomenological extension of HBCHPT, one also counts the nucleon-delta
mass splitting as an additional small parameter and collectively denotes
all small parameters as $\epsilon$. The corresponding expansion, which
also has a consistent power counting, is called the ``small scale
expansion'' (SSE) (because it differs from a chiral expansion due to
the non--vanishing of the $N\Delta$ mass splitting in the chiral limit).\footnote{It is important
to note that the terms of order order $\epsilon^3$ in the SSE that are proportional to the
delta--nucleon mass splitting
$m_\Delta -m_N$ appear at order $p^4$ in the chiral expansion.}
Another important conclusion of
\cite{hhkk} was
that any HBCHPT calculation that wants to calculate $\gamma_2$ and $\gamma_4$ would
have
to be extended to ${\cal O}(p^5)$ before it can incorporate the large
$\Delta$(1232)
related corrections found already at ${\cal O}(\epsilon^3)$
in.\cite{hhkk} Recently, two ${\cal O}(p^4)$ SU(2)
HBCHPT
calculations \cite{Ji,Birse} of polarized Compton scattering in the forward
direction
appeared, from which one can extract one particular
linear combination\footnote{$\gamma_0$ can also be calculated from the
absorption
cross sections of polarized photons on polarized nucleons via the GGT 
sum rule,\cite{GGT} as pointed out in.\cite{BKKM} In the absence of such data several
groups have tried to extract the required cross sections via a partial wave analysis of
unpolarized absorption cross sections. Recent results of these efforts 
are given in table 2.}
 of three of the four $\gamma_i$, usually called $\gamma_0$:
\begin{eqnarray}
\gamma_0=\gamma_1-\left(\gamma_2+2\gamma_4\right)\cos\theta|_{\theta\rightarrow
0} \;.
\end{eqnarray}
The authors of \cite{Ji,Birse} claimed to have found a huge correction to
$\gamma_0$
at ${\cal O}(p^4)$ relative to the ${\cal O}(p^3)$ result already found in,\cite{BKKM}
casting doubt on the usefulness/convergence of HBCHPT for
spin-polarizabilities. Given
that $\gamma_0$ involves the very two polarizabilities
$\gamma_2,\gamma_4$, the (known) poor convergence for $\gamma_0$
found in \cite{Ji,Birse} should not have come as a surprise. We will come back
to this
point later.

\section{Invariant amplitudes and polarizabilities}
We first want to comment on the extraction of polarizabilities from nucleon
Compton
scattering amplitudes.
In previous analyses \cite{hhkk,bkmrev}
it has always been stated that in order to obtain the spin-polarizabilities
from the
calculated Compton amplitudes, one only has to subtract off
the nucleon tree-level (Born) graphs from the fully
calculated amplitudes. The remainder
in each (spin-amplitude) then started with a factor of $\omega^3$ and the
associated
Taylor-coefficient was related to the spin-polarizabilities. Due to the
(relatively)
simple structure of the spin-amplitudes at this order,
this prescription gives the correct result in the
${\cal O}(p^3)$ HBCHPT \cite{bkmrev}
and the ${\cal O}(\epsilon^3)$ SSE \cite{hhkk} calculations. However, at
${\cal O}(p^4)$ (and also at ${\cal O}(\epsilon^4)$ \cite{ghm}) one has to
resort to a
definition of the (spin-) polarizabilities that is soundly based on
field theory,
in order to make sure that one
only picks up those contributions at $\omega^3$ that are really connected with
(spin-) polarizabilities. In fact, at ${\cal O}(p^4)$
(${\cal O}(\epsilon^4)$) the prescription given in \cite{hhkk,bkmrev} leads
to an
admixture of effects resulting from two successive, uncorrelated $\gamma NN$
interactions with a one nucleon intermediate state. In order to avoid these
problems we
advocate the following
definition for the {\em spin-dependent} polarizabilities in (chiral) effective
field theories:~{\em Given a complete set of spin-structure amplitudes
for Compton scattering to
a certain order in perturbation theory, one first removes all
one-particle (i.e. one-nucleon or one-pion) 
reducible (1PR) contributions from the full spin-structure amplitudes.}
To be more precise, at order ${\cal O}(p^4)$ one removes
$F(\omega)/\omega$ terms from the amplitude, where $F(\omega)$ denotes
the energy dependence of the $\gamma NN$ vertex function.
This prescription has been challenged, see the contribution of 
McGovern to these proceedings.\cite{JMG} I will come back to this later.
Specifically, starting from the general form of the T-matrix for real Compton
scattering
assuming invariance under parity, charge conjugation and time reversal
symmetry, we utilize the following six structure amplitudes $A_i(\omega ,
\theta )$ \cite{hhkk,bkmrev} in the Coulomb gauge,
$\epsilon_0=\epsilon_0^\prime=0$,
\begin{eqnarray}
T &=& A_1(\omega,\theta)\vec{\epsilon}^{\, * \prime}\cdot\vec{\epsilon}
+A_2(\omega,\theta)\vec{\epsilon}^{\, * \prime}\cdot\hat{k} \; \vec{\epsilon}
\cdot\hat{k}^\prime \nonumber\\
&+&iA_3(\omega,\theta)\vec{\sigma}\cdot(\vec{\epsilon}^{\, * \prime}\times
\vec{\epsilon}) \nonumber \\
&+&iA_4(\omega,\theta)\vec{\sigma}\cdot(\hat{k}^\prime \times\hat{k})
\vec{\epsilon}^{\, * \prime} \cdot\vec{\epsilon} \nonumber\\
&+& iA_5(\omega,\theta)\vec{\sigma}\cdot[(\vec{\epsilon}^{\, * \prime} \times
\hat{k}) \vec{\epsilon}\cdot\hat{k}^\prime -(\vec{\epsilon}\times
\hat{k}^\prime ) \vec{\epsilon}^{\, * \prime} \cdot\hat{k}]\nonumber\\
&+& iA_6(\omega,\theta)\vec{\sigma}\cdot[(\vec{\epsilon}^{\, * \prime}\times
\hat{k}^\prime ) \hat{\epsilon}\cdot\hat{k}^\prime -(\vec{\epsilon}\times
\hat{k})\vec{\epsilon}^{\, * \prime} \cdot\hat{k}],
\end{eqnarray}
where $\theta$ corresponds to the c.m. scattering angle,
$\vec{\epsilon},\hat{k}\;
(\vec{\epsilon}^{\, \prime} ,\hat{k}^\prime )$ denote
the polarization vector, direction  of the incident (final) photon while
$\vec{\sigma}$ represents the (spin) polarization vector of the nucleon.
Each (spin-)structure amplitude is now separated into 1PR contributions and a
remainder,
that contains the response of the nucleon's excitation structure to two
photons:
\begin{eqnarray}
A_i(\omega,\theta)=A_i(\omega,\theta)^{\rm 1PR}+
A_i(\omega,\theta)^{\rm exc.} \, ,
i=3,\dots,6\, . \label{separation}
\end{eqnarray}
Taylor-expanding the spin-dependent
$A_i(\omega,\theta)^{\rm 1PR}$ for the case of a proton target in the c.m. frame
into
a power series in $\omega$, the leading terms are linear in $\omega$ and are
given by the venerable low--energy theorems (LETs) 
of Low, Gell-Mann and Goldberger:\cite{low}
\begin{eqnarray}
A_3(\omega,\theta)^{\rm 1PR}&=&\frac{\left[1+2\kappa^{(p)}-(1+\kappa^{(p)})^2\cos
\theta\right]e^2}{2M_{N}^2}\,\omega+{\cal O}(\omega^2),\nonumber\\
A_4(\omega,\theta)^{\rm 1PR}&=&-{(1+\kappa^{(p)})^2e^2\over 2M_{N}^2}\,\omega
+{\cal O}(\omega^2),\nonumber\\
A_5(\omega,\theta)^{\rm 1PR}&=&{(1+\kappa^{(p)})^2e^2\over 2M_{N}^2}\,\omega
+{\cal O}(\omega^2),\nonumber\\
A_6(\omega,\theta)^{\rm 1PR}&=&-{(1+\kappa^{(p)})e^2\over 2M_{N}^2}\,\omega
+{\cal O}(\omega^2)~.
\end{eqnarray}
While it is not advisable to really perform this Taylor-expansion for the
spin-dependent $A_i(\omega,\theta)^{\rm 1PR}$
due to the complex pole structure, one can do so without problems
for the $A_i(\omega,\theta)^{\rm exc.}$ as long as $\omega \ll M_\pi$.
For the case of a proton one then finds
\begin{eqnarray}
A_3(\omega,\theta)^{\rm exc.}&=&4\pi \left[ \gamma_{1}^{(p)}-(\gamma_{2}^{(p)}+
2\gamma_{4}^{(p)}) \cos \theta \right]\omega^3+{\cal O}(\omega^4),\nonumber\\
A_4(\omega,\theta)^{\rm exc.}&=&4\pi\gamma_{2}^{(p)}\omega^3+{\cal
O}(\omega^4),\nonumber\\
A_5(\omega,\theta)^{\rm exc.}&=&4\pi\gamma_{4}^{(p)}\omega^3+{\cal
O}(\omega^4),\nonumber\\
A_6(\omega,\theta)^{\rm exc.}&=&4\pi\gamma_{3}^{(p)}\omega^3 +{\cal O}(\omega^4)
\label{xxx}\; .
\end{eqnarray}
We therefore take Eq.(\ref{xxx}) as starting point for the calculation of the
spin-polarizabilities, which are related to the $\omega^3$ Taylor-coefficients
of
$A_i(\omega,\theta)^{\rm exc.}$.
As noted above, both the ${\cal O}(p^3)$ HBCHPT \cite{bkmrev}
and the ${\cal O}(\epsilon^3)$ SSE \cite{hhkk} results are consistent with this
definition.

\section{Isoscalar polarizabilities}

Utilizing Eqs.(\ref{separation},\ref{xxx}) we
have calculated the first subleading correction, ${\cal O}(p^4)$, to
the four isoscalar spin-polarizabilities $\gamma_i^{(s)}$ already determined to
${\cal O}(p^3)$
in~\cite{bkmrev} in SU(2) HBCHPT.
We employ here the convention \cite{hhkk}
\begin{eqnarray}
\gamma_i^{(p)}=\gamma_i^{(s)}+\gamma_i^{(v)} \; ;\quad
\gamma_i^{(n)}=\gamma_i^{(s)}-\gamma_i^{(v)} \; .
\end{eqnarray}
Contrary to popular opinion we show, that even
at subleading order all four spin-polarizabilities can be given in closed form
expressions
which are free of any unknown chiral counterterms! The only parameters
appearing
in the results are the axial-vector nucleon coupling constant, 
the pion decay
constant $F_\pi$, the pion mass $M_\pi$, 
the mass of the nucleon $m_N$ as
well as
its isoscalar, $\kappa^{(s)} = -0.12$, and isovector, $\kappa^{(v)}=3.7$, 
anomalous magnetic moments.
All ${\cal O}(p^4)$ corrections arise from 25 one-loop $\pi N$ continuum
diagrams, with
the relevant vertices obtained from the well-known SU(2) HBCHPT ${\cal O}(p)$
and
${\cal O}(p^2)$ Lagrangians given in detail in ref.\cite{bkmrev}
To ${\cal O}(p^4)$ we find
\begin{eqnarray}
\gamma_1^{(s)}&=& + \; {e^2g_A^2\over 96\pi^3F_\pi^2M_\pi^2} 
                       \left[1-\mu \,\pi\right]~,\\
\gamma_2^{(s)}&=& + \; {e^2g_A^2\over 192\pi^3F_\pi^2M_\pi^2}  
                       \left[1+\mu \,\frac{(-6+\kappa^{(v)})\pi}{4}\right]~,\\
\gamma_3^{(s)}&=& + \; {e^2g_A^2\over 384\pi^3F_\pi^2M_\pi^2}  
                       \left[1-\mu \, \pi\right]~,\\
\gamma_4^{(s)}&=& - \; {e^2g_A^2\over 384\pi^3F_\pi^2M_\pi^2}  
                       \left[1-\mu \,\frac{11}{4}\pi\right]~, 
\end{eqnarray}
with $\mu = M_\pi / m_N \simeq 1/7$ and the
the numerical values given in table 1. 
\begin{table}[t]
\caption{\label{t1}
Predictions for the spin-polarizabilities in HBCHPT in comparison with the
dispersion
analyses of refs.\protect \cite{Krein,gorchtein,bab} (Mainz1,Mainz2,BGLMN) and
the
${\cal O}(\epsilon^3)$ results of the
small scale expansion \protect \cite{hhkk} (SSE1).
All results are given in the units of $10^{-4}\;{\rm fm}^4$.}
\begin{center}
\begin{tabular}{|c||cc|c||cccc|}
\hline
$\gamma_{i}^{(N)}$ & ${\cal O}(p^3)$ & ${\cal O}(p^4)$ & Sum  & Mainz1 & Mainz2
&
BGLMN & SSE1  \\
\hline
$\gamma_{1}^{(s)}$ & $+ 4.6$ & $-2.1$ & $+2.5$ & $+5.6$ &$+5.7$ &  $+4.7$
&$+4.4$\\
$\gamma_{2}^{(s)}$ & $+ 2.3$ & $-0.6$ & $+1.7$ & $-1.0$ &$-0.7$ &  $-0.9$
&$-0.4$\\
$\gamma_{3}^{(s)}$ & $+ 1.1$ & $-0.5$ & $+0.6$ & $-0.6$ &$-0.5$ &  $-0.2$
&$+1.0$\\
$\gamma_{4}^{(s)}$ & $- 1.1$ & $+1.5$ & $+0.4$ & $+3.4$ &$+3.4$ &  $+3.3$
&$+1.4$\\
\hline
$\gamma_{1}^{(v)}$ & - & $-1.3$ & $-1.3$ & $-0.5$ & $-1.3$ & $-1.6$ & - \\
$\gamma_{2}^{(v)}$ & - & $-0.2$ & $-0.2$ & $-0.2$ & $+0.0$ & $+0.1$ & - \\
$\gamma_{3}^{(v)}$ & - & $+0.1$ & $+0.1$ & $-0.0$ & $+0.5$ & $+0.5$ & - \\
$\gamma_{4}^{(v)}$ & - & $+0.0$ & $+0.0$ & $+0.0$ & $-0.5$ & $-0.6$ & - \\
\hline
\end{tabular}
\end{center}
\end{table}
\noindent
The leading $1/M_\pi^2$ behavior
of the
isoscalar spin-polarizabilities is not touched by the ${\cal O}(p^4)$
correction, as expected.
With the notable exception of $\gamma_4^{(s)}$,
which even changes its sign due to a large ${\cal O}(p^4)$ correction, we show
that
this first subleading order of $\gamma_1^{(s)},\gamma_2^{(s)},\gamma_3^{(s)}$
amounts
to a 25-45\% correction to the leading order result. This does not quite
correspond
to the expected $M_\pi/m_N$ correction of (naive) dimensional analysis, but can
be considered acceptable. 
The large correction in $\gamma_4^{(s)}$
should be considered accidental. It is not related to the
the large $\Delta$ effects found in the SSE calculation of,\cite{hhkk} 
because these will only show up at ${\cal O}(p^5)$ in the HBCHPT
framework. This can easily be understood: When the delta is not an active dof
in the effective field theory, it can not modify the leading singularity of order
$1/M_\pi^2$ due to decoupling. In principle, the spin--dependent local operator
of dimension two $\sim c_4$ inserted in one loop graphs could modify the
$1/M_\pi$ terms. However, such contributions cancel in the sum of all loop
graphs. This is different from the from the spin--independent case, where
a particular combination of certain $c_i$ generates an important pion loop
correction at fourth order.

\section{Isovector polarizabilities}

We further report the first results for the four {\em isovector}
spin-polarizabilities
$\gamma_i^{(v)}$ obtained in the framework of chiral effective field theories.
Previous calculations at ${\cal O}(p^3)$ \cite{bkmrev} and
${\cal O}(\epsilon^3)$ \cite{hhkk} were only sensitive to the isoscalar
spin-polarizabilities $\gamma_i^{(s)}$, therefore this calculation gives the
first indication from a 
chiral effective field theory about the magnitude of the difference
in the low-energy spin structure between proton and neutron. As in the case of
the isoscalar spin-polarizabilities there are again no unknown counterterm
contributions
to this order in the $\gamma_i^{(v)}$. All ${\cal O}(p^4)$
contributions arise from 16 one-loop
$\pi N$ continuum diagrams with the relevant ${\cal O}(p),\;{\cal O}(p^2)$
vertices again obtained from the Lagrangians given in ref.\cite{bkmrev}
To ${\cal O}(p^4)$ one finds
\begin{eqnarray}
\gamma_1^{(v)}&=&  {e^2g_A^2\over 96\pi^3F_\pi^2M_\pi^2} 
                       \left[0-\mu \, \frac{5\pi}{8}\right]~,\\
\gamma_2^{(v)}&=&  {e^2g_A^2\over 192\pi^3F_\pi^2M_\pi^2}  
                       \left[0-\mu \, \frac{(1+\kappa^{(s)})\pi}{4}\right]~,\\
\gamma_3^{(v)}&=&  {e^2g_A^2\over 384\pi^3F_\pi^2M_\pi^2}  
                       \left[0+\mu \, \frac{\pi}{4}\right]~,\\
\gamma_4^{(v)}&=&   0~, 
\end{eqnarray}
with the numerical values again given in table 1.
The result of our investigation is
that the size of the $\gamma_i^{(v)}$ really tends to be an order of
magnitude smaller than the one of the $\gamma_i^{(s)}$ (with the possible
exception
of $\gamma_1^{(v)}$), supporting the scaling expectation, $\gamma_i^{(v)}\sim
(M_\pi/m_N)
\gamma_i^{(s)}$ from (naive) dimensional analysis. This is reminiscent of the
situation
in the spin-independent electromagnetic polarizabilities
$\bar{\alpha}^{(v)},\bar{\beta}^{(v)}$,\cite{ulfco} which are also
suppressed
by one chiral power relative to their isoscalar partners
$\bar{\alpha}^{(s)},\bar{\beta}^{(s)}$.

\section{Forward and backward polarizabilities}
Finally, we want to comment on the comparison between our results and
existing
calculations using dispersion analyses. Given our comments on the convergence
of the
chiral expansion for the (isoscalar) spin-polarizabilities,\cite{hhkk} we
reiterate
that we do not believe our ${\cal O}(p^4)$ HBCHPT result for
$\gamma_2^{(s)},\gamma_4^{(s)}$
to be meaningful. Their large inherent $\Delta$(1232) related contribution just
cannot be
included (via a counterterm) before ${\cal O}(p^5)$ in HBCHPT that only deals
with pion
and nucleon degrees of freedom. In table 1 it is therefore
interesting to note that by adding (``by hand'') the delta-pole contribution of
$\sim-2.5\cdot 10^{-4}$fm$^4$ found in \cite{hhkk} to $\gamma_2^{(s)}$ one could
get quite
close to the range for this spin-polarizability
as suggested by the dispersion analyses.\cite{Krein,gorchtein,bab} Similarly,
adding
$\sim+2.5\cdot 10^{-4}$fm$^4$ to $\gamma_4^{(s)}$ as suggested by~\cite{hhkk} also
leads
quite close to the range advocated by the dispersion results.\cite{Krein,gorchtein,bab}
However, such a procedure is of course not legitimate in an effective field
theory,
but it raises the hope that an extension of the ${\cal O}(\epsilon^3)$ SSE
calculation of
\cite{hhkk} that includes explicit delta degrees of freedom could lead to a
much
better behaved perturbative expansion for the isoscalar spin-polarizabilities.
Whether
this expectation holds true will be known quite soon.\cite{ghm}
For the isovector spin-polarizabilities we have given the first predictions
available
from effective field theory. In general the agreement with the range advocated
by the
dispersion analyses is quite good. 

\medskip\noindent
Furthermore,
in table 2 we give a comparison of our results for those linear combinations of
the
$\gamma_i$
that typically are the main focus of attention in the literature. However, we
re-emphasize
that we do not consider our ${\cal O}(p^4)$ HBCHPT predictions for
$\gamma_0^{(s)}, \gamma_\pi^{(s)}$ to be meaningful, because they involve
$\gamma_2^{(s)},\gamma_4^{(s)}$. 
The corresponding isovector combinations,
however, again
seem to agree quite well with the dispersive results and so far we have no
reason
to suspect that they might be affected by the poor convergence behavior of
some of their
isoscalar counterparts. We further note that our ${\cal O}(p^4)$ HBCHPT
predictions for $\gamma_0^{(s,v)}$ differ from the ones given in two recent
calculations.\cite{Ji,Birse} 
As noted above this difference solely arises from a different definition
of nucleon spin-polarizabilities. If we (``by hand'')
Taylor-expand our $\gamma NN$ vertex
functions $F(\omega)$ in powers of $\omega$ and include the resulting terms into the the
$\gamma_0$
structure, we obtain the ${\cal O}(p^4)$ corrections $\gamma_0^{(s)}=-6.9,\;
\gamma_0^{(v)}=-1.6$ in units of $10^{-4}$fm$^4$, in numerical (and analytical)
agreement
with.\cite{Ji,Birse} This brings us to an important point: Once the first
polarized
Compton asymmetries have been measured, it has to be checked very carefully
whether the
same input data fitted to the terms we define as 1PR plus the additional free
$\gamma_i$
parameters leads to the same numerical fit-results for the
spin-polarizabilities as
in the dispersion theoretical codes usually employed to extract polarizabilities
from
Compton data. Small
differences for example in the treatment of the pion/nucleon pole could lead to
quite
large systematic errors in the determination of the $\gamma_i$. Such studies
are under
way.\cite{ghm} Also, the controversy about how to subtract the 1PR pieces
in a non--relativistic approach like HBCHPT can be setteld employing a 
Lorentz--invariant formulation of baryon CHPT~\cite{BL} since in that
context the subtraction is unambigouos and the heavy fermion limit can 
be obtained easily.
\begin{table}[t]
\caption{\label{t2}
Predictions for the so-called forward (backward) spin-polarizabilities
$\gamma_0$
($\gamma_\pi$). For a definition of units and references see table 1.}
\begin{center}
\begin{tabular}{|c||cc|c||cccc|}
\hline
$\gamma_{i}^{(N)}$ & ${\cal O}(p^3)$ & ${\cal O}(p^4)$ & Sum  & Mainz1 & Mainz2
&
BGLMN & SSE1  \\
\hline
$\gamma_{0}^{(s)}$   & $+ 4.6$ & $-4.5$ & $+0.1$ & $-0.2$  &$-0.4$  & $-1.0$&
+2.0 \\
$\gamma_{0}^{(v)}$   & -       & $-1.1$ & $-1.1$ & $-0.3$  &$-0.4$  & $-0.5$& -
\\
$\gamma_{\pi}^{(s)}$ & $+ 4.6$ & $+0.3$ & $+4.9$ & $+11.4$ &$+11.8$ & $+10.4$&
+6.8\\
$\gamma_{\pi}^{(v)}$ & -       & $-1.5$ & $-1.5$ & $-0.7$  &$-2.4$  &  $-2.7$&
- \\ \hline
\end{tabular}
\end{center}
\end{table}
\section*{Acknowledgments}
These results have been obtained in collaboration with George Gellas and
Thomas Hemmert, to whom I express my sincere gratitude. I also would like
to thank the organizers for putting together such an interesting program.


\begin{thebibliography}{99}
\bibitem{1} F.J. Federspiel et al., Phys. Rev. Lett. {\bf 67} (1991) 1511;
E.L. Hallin et al., Phys. Rev. {\bf C48} (1993) 1497; 
A. Zieger et al., Phys. Lett. {\bf B278} (1992) 34; 
B.E. MacGibbon et al., Phys. Rev. {\bf C52} (1995) 2097.
\bibitem{BKMpola} V. Bernard, N. Kaiser and Ulf-G. Mei{\ss}ner,
Phys. Rev. Lett. {\bf 67} (1991) 1515.
\bibitem{ulfco} V. Bernard, N. Kaiser, A. Schmidt and Ulf-G. Mei{\ss}ner,
Phys. Lett. {\bf B319} (1993) 269; Z. Phys. {\bf A348} (1994) 317.
\bibitem{ragusa} See e.g. S. Ragusa, Phys. Rev. {\bf D47}, 3757 (1993);
{\bf D49} (1994) 3157.
\bibitem{trhproc}T.R. Hemmert, ``Theory of the nucleon spin--polarizabilities~II'',
 {\it these proceedings}.
\bibitem{bab} D. Babusci et. al., Phys. Rev. {\bf C58}  (1998)  1013.
\bibitem{gorchtein} D. Drechsel, M. Gorchtein, B. Pasquini and M.
Vanderhaeghen,
Phys. Rev. {\bf C61} (2000) 015204.
\bibitem{legs} J.I. Tonnison et. al., Phys. Rev. Lett. {\bf 80} (1998) 4382.
\bibitem{Wiss}F. Wissmann et al., Nucl. Phys. {\bf A 660} (1999) 232.
\bibitem{hhkk} T.R. Hemmert, B.R. Holstein, J. Kambor and G. Kn{\"o}chlein,
Phys. Rev. {\bf D57} (1998) 5746.
\bibitem{bkmrev} V. Bernard, N. Kaiser and Ulf-G. Mei{\ss}ner, Int. J. Mod.
Phys. {\bf E4} (1995) 193.
\bibitem{Ji}X. Ji, C.-W. Kao and J. Osborne, Phys. Rev. {\bf D61} (2000)
  074003.
\bibitem{Birse} K.B. Vijaya Kumar, J.A. McGovern and M.C. Birse,
{\tt hep-ph/9909442}. 
\bibitem{GHMprl}G.C. Gellas, T.R. Hemmert and Ulf-G. Mei{\ss}ner,
Phys. Rev. Lett. {\bf 85} (2000) 14.
\bibitem{Birse2} K. B. Vijaya Kumar, J.A. McGovern and  M.C. Birse,
Phys. Lett. {\bf B479} (2000) 167.
\bibitem{BKKM} V. Bernard, N. Kaiser, J. Kambor and Ulf-G. Mei\ss ner, Nucl.
Phys. {\bf B388} (1992) 315.
\bibitem{SSE} T.R. Hemmert, B.R. Holstein and J. Kambor, Phys. Lett. {\bf
B395} (1997) 89; J. Phys. {\bf G24} (1998) 1831.
\bibitem{GGT} M. Gell-Mann, M.L. Goldberger and W.E. Thirring, Phys. Rev.
{\bf 95} (1954) 1612.
\bibitem{ghm} G.C. Gellas, T.R. Hemmert and Ulf-G. Mei{\ss}ner, {\em forthcoming}.
\bibitem{JMG}J.A. McGovern, M.C. Birse and K.B. Vijaya Kumar, {\tt nucl-th/0007015}. 
\bibitem{low} M. Gell-Mann and M.L. Goldberger, Phys. Rev. {\bf 96},
(1954) 1433; F.E. Low, Phys. Rev. {\bf 96} (1954).
\bibitem{Krein} D. Drechsel, G. Krein and O. Hanstein, Phys. Lett. {\bf B420}
(1998) 248.
\bibitem{BL} T. Becher and H. Leutwyler, Eur. Phys. J. {\bf C9} (1999) 643.
\end{thebibliography}
\end{document}